\documentclass[global,twocolumn]{svjour}
\usepackage{latexsym}
\usepackage{graphics}
\journalname{Applied physics B}
\begin{document}
\title{1D Bose Gases in an Optical Lattice}
\author{Michael K\"ohl, Thilo St{\"o}ferle, Henning Moritz, Christian Schori
\and Tilman Esslinger}

\institute{Institute for Quantum Electronics, ETH Z\"urich
H\"onggerberg, CH-8093 Z\"urich, Switzerland}

\date{Received: \today / Revised version: date}

\maketitle

\begin{abstract}
We report on the study of the momentum distribution of a
one-dimensional Bose gas in an optical lattice. From the momentum
distribution we extract the condensed fraction of the gas and
thereby measure the depletion of the condensate and compare it
with a theorical estimate. We have measured the coherence length
of the gas for systems with average occupation $\bar{n}>1$ and
$\bar{n}<1$ per lattice site.
\end{abstract}

\section{The one-dimensional Bose gas in an optical lattice}
A one-dimensional gas can be created in a trap when the confining
potential restricts the motion of the particles to one dimension
with the transverse motional degrees of freedom being frozen out.
A cigar shaped harmonic trapping geometry is characterized by the
frequencies $\omega_\perp$ in the two strongly confining axes and
$\omega_{ax}$ in the weakly confining axis. A kinematically
one-dimensional situation is achieved when all particles occupy
only the ground state in the radial directions, which implies that
both the thermal energy $k_B T$ and the interaction energy $\mu$
have to be much smaller than the transverse energy level spacing.
In general, one-dimensional systems exhibit increased quantum
fluctuations of the phase, such that for a homogeneous 1D gas
Bose-Einstein condensation is prevented. For {\it trapped}
low-dimensional gases the cross-over to a Bose-Einstein condensate
takes place at a finite temperature $k_B T_{1D}=N \hbar
\omega_{ax}/\ln(2N)$, where $N$ is the number of particles in the
1D system \cite{Ketterle1996,Petrov2000a}. The fluctuating phase
alters the properties of the gas
\cite{Kuehner1998,Batrouni2002,Kollath2004,Wessel2004,Stoferle2004}.
One-dimensional quantum systems exhibit a wealth of fascinating
phenomena whose explanations go beyond the mean-field description
\cite{Giamarchi2004}.

One-dimensional trapped Bose-Einstein condensates were recently
created and studied \cite{Moritz2003} using an optical lattice
consisting of two mutually perpendicular standing wave laser
fields. In this geometry the optical lattice forms an array of
one-dimensional tubes, each filled with a Bose-Einstein
condensate. This experiment revealed the distinctively different
excitation spectrum of a one-dimensional quantum system as
compared to its three dimensional counterpart \cite{Menotti2002}.
In a previous experiment a Bose condensates was loaded into a
two-dimensional optical lattices to study the coherence between
the tubes. In that experiment the tunnel-coupling between adjacent
tubes was larger than the axial oscillation frequency, thereby an
array of strongly coupled tubes was created \cite{Greiner2001b}.
Very recently the lifetime of one-dimensional gases created in an
optical lattice were studied \cite{Laburthe2004}. In elongated
magnetic and optical traps a regime was accessed where a Bose
condensate with $\mu \leq \hbar \omega_\perp$ coexisted with a
three-dimensional thermal cloud \cite{Gorlitz2001,Schreck2001}. In
similar elongated traps studies of solitons \cite{Strecker2002}
and of enhanced phase fluctuations have been performed
\cite{Dettmer2001,Richard2003}.

When the one-dimensional Bose gas is exposed to an additional
optical lattice potential in axial direction the bosons may become
localized in the minima of a periodic potential and the system can
then be described by the Bose-Hubbard Hamiltonian
\cite{Fisher1989,Jaksch1998}:
\begin{eqnarray}
H=-\widetilde{J} \sum_{ \langle i,j \rangle} \hat{a}^\dagger_{j}
\hat{a}_i+ \sum_i \epsilon_i n_i+\frac{U}{2} \sum_i n_i (n_i-1).
\end{eqnarray}
$\widetilde{J}$ denotes the hopping matrix element between
neighboring lattice sites and determines the rate of which a
particle disappears from lattice site $i$ and tunnels to the
adjacent lattice site $j$ ($\hat{a}_i$ and $\hat{a}^\dagger_i$ are
the annihilation and creation operators for an atom at lattice
site $i$, respectively). The total tunnel coupling is $J=z
\widetilde{J}$ with $z$ being the coordination number of the
lattice. The inhomogenity of the atom trap is characterized by
$\epsilon_i=\frac{m}{2}(i\omega_{ax}d)^2$ where $m$ is the atomic
mass and $d$ the lattice spacing. The occupation number of lattice
site $i$ is denoted $n_i$ and $U$ is the onsite-interaction energy
between two bosons on the same lattice site. The one-dimensional
Bose-Hubbard Hamiltonian exhibits a transition from a superfluid
phase to a Mott insulating phase for a ratio $(U/J)_c\simeq 2$
\cite{Kuehner1998,Batrouni2002,Kollath2004,Wessel2004}, which we
experimentally demonstrated recently \cite{Stoferle2004}. Due to
stronger quantum fluctuations of the phase in a one-dimensional
quantum system this value is lower than the corresponding critical
value in three dimensions $(U/J)_c=5.8$
\cite{Fisher1989,Jaksch1998,Greiner2002}. Tuning the depth of the
periodic potential changes the parameter $U/J$, which leads to
increased effects of interparticle interactions and small filling
of the lattice sites may turn the system into a gas of hardcore
bosons \cite{Cazallila2003,Paredes2004}. In the inhomogeneous
system Mott insulating regions with commensurate filling coexist
with superfluid regions with incommensurate filling and the
insulator is formed in a crossover type transition
\cite{Batrouni2002}.

\section{Experimental setup}
\subsection{Bose condensates in an optical lattice}
In the experiment, we collect up to $2 \times 10^9$ $^{87}$Rb
atoms in a vapor cell magneto-optical trap. After polarization
gradient cooling and optical pumping into the $|F=2, m_F=2\rangle$
hyperfine ground state the atoms are captured in a magnetic
quadrupole trap. After magnetic transport of the trapped atoms
over a distance of 40\,cm the magnetic trapping potential is
converted into the harmonic and elongated potential of a QUIC trap
\cite{Esslinger1998}. Subsequently, we perform radio frequency
induced evaporation of the cloud over a period of 25\,s. After
evaporation we observe almost pure three-dimensional Bose-Einstein
condensates of up to $1.5 (0.2)\times 10^5$ atoms. Following the
condensation we adiabatically change the trapping geometry to an
approximately spherical symmetry with trapping frequencies of
$\omega_x=2 \pi \times 18$\,Hz, $\omega_y=2 \pi \times 20$\,Hz,
and $\omega_z=2 \pi \times 22$\,Hz. This reduces the peak density
by a factor of 4 and allows us to load the optical lattice more
uniformly.

The optical lattice is formed by three retro-reflected laser
beams. Each beam is derived from a laser diode at a wavelength of
$\lambda=826$\,nm. At the position of the condensate the beams are
circularly focused to $1/e^2$-radii of $120$\,$\mu\textrm{m}$ ($x$
and $y$ axes) and $105$\,$\mu\textrm{m}$ ($z$). The three beams
possess mutually orthogonal polarizations and their frequencies
are offset with respect to each other by several ten MHz. We
stabilize the lasers to a high-finesse Fabry-Perot cavity, thereby
reducing their line width to $\sim 10\,\textrm{kHz}$. In order to
load the condensate into the ground state of the optical lattice,
the intensities of the lasers are slowly increased to their final
values using an exponential ramp with a time constant of
$25\,\textrm{ms}$ and a duration of $100\,\textrm{ms}$. The
resulting optical potential depths $V_{x,y,z}$ are proportional to
the laser intensities and are conveniently expressed in terms of
the recoil energy $E_R=\frac{\hbar^2 k^2}{2 m}$ with $k=\frac{2
\pi}{\lambda}$.

\subsection{Preparation of one-dimensional quantum gases}
Using the optical lattice we realize one-dimensional quantum gases
ordered in an array of one-dimensional tubes. Two lattice axes are
ramped to a fixed value $V_\perp \equiv V_x=V_z=30\,E_R$. In this
configuration the transverse tunnelling rates $J_x$ and $J_z$ are
small compared to the duration of the experiment. We therefore
neglect tunnelling between neighboring tubes and assume an array
of isolated 1D gases. An additional optical lattice is applied
along the symmetry axis with a potential depth $V_{ax} \equiv V_y
\ll V_\perp$. The potential depth $V_{ax}$ controls the value of
$U/J$.

\section{Results}

\subsection{Momentum distribution}
We have studied the momentum distribution of the gas by imaging
the expanding atom cloud after being released from the optical
lattice. Prior to switching off the lattice the laser intensity of
the axial lattice is increased to $\simeq 25 E_{R}$ for
40\,$\mu$s. Then all laser beams are extinguished simultaneously
within 5\,$\mu$s and the magnetic trapping potential is switched
off within 300\,$\mu$s. This procedure may be regarded as a
projection of the wave function of the atoms onto the Bloch
momentum states for a lattice depth of 25\,$E_{R}$. This method
offers the advantage of controlling adiabatic processes during the
switch-off of the lattice. These might affect the momentum
distribution of the released atoms \cite{Paredes2004}.

For our one-dimensional gases the collisional interaction during
expansion -- which would give rise to a Thomas-Fermi type envelope
of the density distribution in the y-direction -- is small. This
is related to the extremely fast radial expansion of the clouds,
which are initially confined to the harmonic oscillator ground
state $a_\perp=\sqrt{\hbar/(m \omega_\perp)}$ with a trapping
frequency of $\omega_\perp=2\pi \times 33$\,kHz. The peak density
decreases proportional to $1/\sqrt{1+(\omega_\perp t)^2}$ which is
much faster than the timescale for axial expansion that is given
by the axial trapping frequency of $\omega_{ax}=2 \pi \times
80$\,Hz. This axial confinement is introduced by the gaussian
waist of the laser beams forming the array of one-dimensional
tubes.

\begin{figure}[htbp]
\resizebox{\columnwidth}{!}{\includegraphics{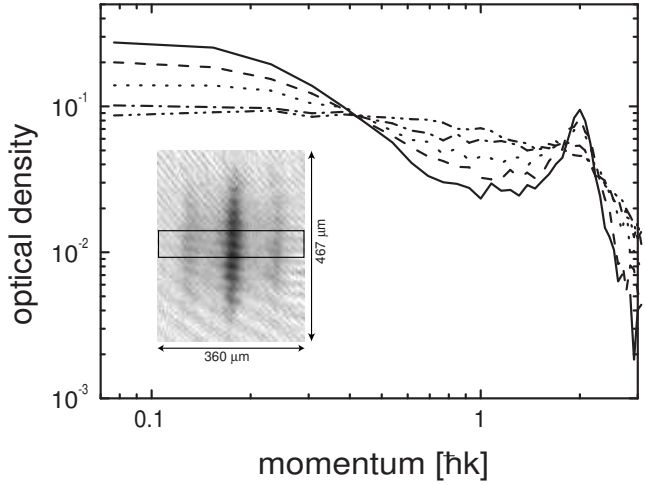}}
\caption{Measured momentum distribution for $N=1.5\times 10^5$
atoms for different depth of the axial lattice potential. Top to
bottom on the left hand site: $V_{ax}=5$, $V_{ax}=7$, $V_{ax}=9$,
$V_{ax}=12$, $V_{ax}=18$ (in units of $E_R$). The inset shows an
absorption image after 10\,ms time of flight where the area of
averaging is indicated.} \label{fig1}
\end{figure}

We have studied the resulting momentum distribution after 15\,ms
of free expansion of the cloud. The atoms were imaged by resonant
absorption imaging and we determine the optical density in the
center of the cloud by averaging over a $62\,\mu$m wide section in
the image. Fig. \ref{fig1} shows the measured momentum
distribution for a sample with an average filling of $\bar{n}=1.2$
atoms per lattice site. We calculate the average filling from the
mean density of atoms which is derived known atom number and the
measured trapping frequencies and includes a modified coupling
strength due to the localization of the atoms in the optical
lattice potential \cite{Kramer2003}. The atom number distribution
at a given lattice site in the superfluid phase can be
approximated by a Poisson distribution. Therefore, even with an
average occupation number of $\bar{n}=0.6$ the probability of
finding doubly or multiply occupied lattice sites is on the order
of $10\%$.

\subsection{Quantum depletion of the condensate}
\begin{figure}[htbp]
\resizebox{\columnwidth}{!}{\includegraphics{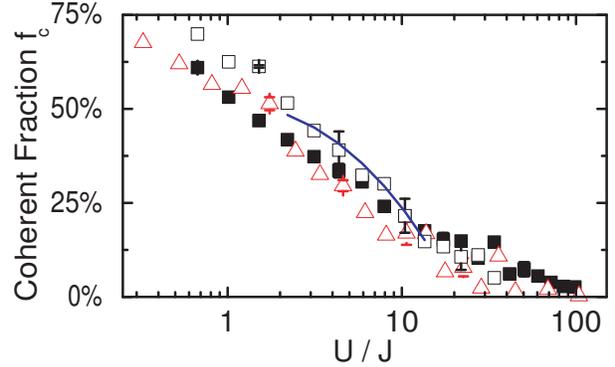}}
\caption{Coherent fraction of the 1D Bose gas as measured from the
time-of-flight image. One-dimensional system with $\bar{n}=1.2$
(solid squares), one-dimensional system with $\bar{n}=0.6$ (open
squares), three-dimensional gas with $\bar{n}=1$ (triangles). Part
of the experimental data are taken from \cite{Stoferle2004}. The
solid line shows the calculated quantum depletion according to
equation (\ref{eq2}) using $N_w = 40$.} \label{fig2}
\end{figure}

When the potential depth of the lattice is raised the strength of
the atom-atom interaction increases. This is accompanied by
quantum depletion of the condensate \cite{Bogoliubov1947}. The
reduced condensate fraction of the system diminishes the contrast
in the matter wave interference pattern after the atoms have been
released from the optical lattice. To extract the number of
coherent atoms $N_{coh}$ from the interference pattern, the
interference peaks at $0\hbar k$, $\pm 2\hbar k$ and $\pm 4 \hbar
k$ are fitted by gaussians. Incoherent atoms, both due to
localization of the atoms in the lattice or due to depletion of
the condensate, give rise to a broad gaussian background centered
at zero quasi momentum which dominates for higher $V_{ax}$
\cite{Hadzibabic2004}. Taking this fit as a measure of the number
of incoherent atoms $N_{incoh}$, we calculate the coherent
fraction $f_c=\frac{N_{coh}}{N_{coh}+N_{incoh}}$
\cite{Stoferle2004}. The measured data are shown in figure
\ref{fig2}. For the one-dimensional Bose gas in an optical lattice
the quantum depletion is calculated by \cite{Kramer2003}
\begin{eqnarray}
\frac{n}{n_0}=\sqrt{\frac{1}{2 \pi^2} \frac{m}{m^*}
\frac{1}{a_{1D} n_{1D}}} \ln\left( \frac{4 N_w}{\pi}\right).
\label{eq2}
\end{eqnarray}
Here $m$ is the atomic mass, $m^*$ is the effective mass in the
lattice, $n_{1D}$ is the 1D density, $a_{1D}$ is the
one-dimensional scattering length  \cite{Olshanii1998} and $N_w$
is the number of potential wells occupied by atoms. This
expression holds in the tight binding regime but for small
depletion. Figure \ref{fig2} displays the measured coherent
fraction together with the calculated values for the quantum
depletion, which show good agreement in the range of validity of
eq.\,(\ref{eq2}).

\subsection{Coherence length}
\begin{figure}[htbp]
\resizebox{\columnwidth}{!}{\includegraphics{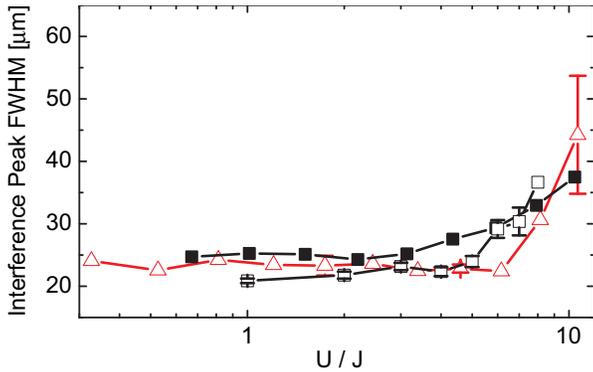}}
\caption{Width of the central momentum peak as a measure of the
phase coherence length of the system. One-dimensional system with
$\bar{n}=1.2$ (solid squares), one-dimensional system with
$\bar{n}=0.6$ (open squares), three-dimensional gas with
$\bar{n}=1$ (triangles). Part of the data are taken from ref.
 \cite{Stoferle2004}.} \label{fig3}
\end{figure}

From the width of the central coherent momentum component the
coherence length of the gas in the optical lattice can be
inferred. A Bose-Einstein condensate exhibits a long coherence
length which leads to a small width of the interference peak. Due
to the inhomogeneity of the trap the transition to the Mott
insulator sets in at those positions, where the local density
becomes commensurable with the spacing of the potential wells in
the optical lattice. This process breaks up the condensed cloud
into smaller units and therefore reduces the coherence length of
the system. In figure \ref{fig3} we show the change of the
coherence length as a function of the parameter $U/J$. For the
one-dimensional gas with an average occupancy of $\bar{n}=1.2$ we
observe a kink in the coherence length at $U/J\simeq 2$ which
indicates the onset of the Mott insulating phase. For the
three-dimensional gas with $\bar{n}=1$ we find the Mott insulator
transition at $U/J \simeq 6$, in agreement with the mean field
theory. For very small one-dimensional systems with $\bar{n}=0.6$
we observe that the coherence length of the sample starts to
increase around $U/J=4$. This is in qualitative agreement with the
theoretical prediction that the Mott-insulator transition occurs
for $U/J\simeq 2$ only for a chemical potential corresponding a
density $\bar{n}=1$ but at larger values for $\bar{n}<1$
\cite{Kuehner1998}.

\section{Conclusion}
We have studied the momentum distribution of one-dimensional Bose
gases in an optical lattice for different average occupation of
the lattice sites. We find that the coherent fraction of atoms is
independent of the atomic density when we tune the interaction
strength $U/J$ from the superfluid to the Mott-insulating regime.
The change of the coherence length appear to indicate that the
system enters the Mott-insulator for small atomic densities only
for larger values of $U/J$.

\bigskip

We acknowledge stimulating discussions with Thierry Giamarchi,
Corinna Kollath and Stefan Wessel and support from SNF and QSIT.

\end{document}